\documentclass[numbers,sort&compress,3p,onecolumn,times]{elsarticle}
\pdfoutput=1
\journal{Results in Physics}
\usepackage{amsmath,amssymb}
\usepackage{xspace}
\usepackage[bookmarksnumbered,bookmarksopen,bookmarksopenlevel=1,colorlinks=false,plainpages=false,pdfpagelabels]{hyperref}
\hypersetup{linkbordercolor = {1 1 1}}
\usepackage{graphicx,graphics}
\usepackage{array}
\usepackage{ifthen}
\usepackage{multirow}
\usepackage{lineno}
\usepackage{rotating}

\usepackage{dcolumn}
\usepackage{longtable}
\newenvironment{acknowledgments}{\section*{Acknowledgments}}{\par}

\wlog{Patching amsLaTeX equation handling to deal with the presence of line numbers}
\usepackage{etoolbox}%
\newcommand*\linenomathpatch[1]{%
\cspreto{#1}{\linenomath}%
\cspreto{#1*}{\linenomath}%
\csappto{end#1}{\endlinenomath}%
\csappto{end#1*}{\endlinenomath}}
\linenomathpatch{equation}
\linenomathpatch{gather}
\linenomathpatch{multline}
\linenomathpatch{align}
\linenomathpatch{alignat}
\linenomathpatch{flalign}
\def\ijazz{\textsc{IJazZ2.0}\xspace}
\def\rl{\ensuremath{{r_{\ell}}}\xspace}
\def\rll{\ensuremath{{r_{\ell\ell}}}\xspace}
\def\sl{\ensuremath{{\sigma_{\ell}}}\xspace}
\def\sll{\ensuremath{{\sigma_{\ell\ell}}}\xspace}
\def\mll{\ensuremath{{m_{\ell\ell}}}\xspace}
\def\gev{\ensuremath{\mathrm{GeV}}\xspace}

\def\zll{\ensuremath{Z\to\ell\ell}\xspace}
\def\dyll{\ensuremath{Z\to\ell\ell}\xspace}
\def\data{\ensuremath{\mathrm{data}}\xspace}
\def\mc{\ensuremath{\mathrm{mc}}\xspace}
\def\nll{\ensuremath{{nll}}\xspace}

\def\ie{\emph{i.e.}\xspace}
\def\eg{\emph{e.g.}\xspace}

\def\nic{\ensuremath{n_{ic}}\xspace}
\def\pic{\ensuremath{p_{ic}}\xspace}
\def\mjc{\ensuremath{m_{jc}}\xspace}
\def\djc{\ensuremath{\delta_{jc}}\xspace}
\def\aijc{\ensuremath{\alpha_{ijc}}\xspace}

\def\tev{\ensuremath{\mathrm{TeV}}\xspace}
\def\gev{\ensuremath{\mathrm{GeV}}\xspace}

\def\vtheta{\ensuremath{\vec{\theta}}\xspace}
\def\vthetast{\ensuremath{{\vtheta}^*}\xspace}

\def\pt{\ensuremath{p_\mathrm{T}}\xspace}
\def\ptbar{\ensuremath{\overline{p}_\mathrm{T}}\xspace}

\def\mup{\ensuremath{\mu^+}\xspace}
\def\mum{\ensuremath{\mu^-}\xspace}
\def\kmm{\ensuremath{z_{\mu\mu}}\xspace}
\def\dyee{\ensuremath{Z\to e^+ e^-}\xspace}
\def\dymmg{\ensuremath{Z\to \mu^+ \mu^- \gamma}\xspace}
\def\dyll{\ensuremath{Z\to\ell\ell}\xspace}
\def\dy{\ensuremath{\mathrm{DY}}}
\def\Mmmg{\ensuremath{m_{\mu\mu\gamma}}\xspace}
\def\Mmm{\ensuremath{m_{\mu\mu}}\xspace}
\def\ptg{\ensuremath{p_{\mathrm{T}\, \gamma}}\xspace}

\def\ptmum{\ensuremath{p_{\mathrm{T}\, \mum}}\xspace}
\def\ptmup{\ensuremath{p_{\mathrm{T}\, \mup}}\xspace}

\def\MZ{\ensuremath{m_Z}\xspace}
\def\Vdy{\ensuremath{{V_{\mathrm{DY}}}}\xspace}
\def\rg{\ensuremath{{r_{\gamma}}}\xspace}
\def\sg{\ensuremath{{\sigma_{\gamma}}}\xspace}
\def\drg{\ensuremath{\delta\rg}\xspace}
\def\drgm{\ensuremath{\overline{\drg}}\xspace}

\begin{document}

\title{Lepton and photon energy scale and resolution corrections based on the minimization of an analytical likelihood: \ijazz}

\date{\today}
\address[saclay]{IRFU, CEA, Université Paris-Saclay, F-91191 Gif-sur-Yvette, France}
\author[saclay]{F. Couderc}
\author[saclay]{P. Gaigne}
\author[saclay]{M. Ö. Sahin}
\begin{abstract}
We present a novel method to determine lepton energy scale and resolution corrections by means of an analytical likelihood maximization applied to Drell--Yan \(Z \to \ell\ell\) events. The approach relies on an exact analytical treatment of the Gaussian energy smearing model, avoiding random-number-based convolution techniques.
This formulation results in a fully differentiable likelihood enabling the use of automatic differentiation algorithms, and thus a substantial reduction in computational cost. The method, implemented in the \ijazz software, allows the simultaneous extraction of scale and resolution parameters across multiple lepton categories defined by detector or kinematic variables. We validate the technique using toy Monte Carlo studies and realistic Pythia-based simulations, demonstrating unbiased parameter recovery and accurate uncertainty estimates. Particular attention is given to categorizations involving lepton transverse momentum, for which a relative-\(p_T\) strategy is introduced to mitigate biases induced by category migration and kinematic correlations.
The method is further adapted to photon-energy scale measurement in \(Z \to \mu^+\mu^-\gamma\) decays. Compared to conventional approaches, the analytical method improves numerical stability, robustness of the minimization, and computational performance, making it well suited for large-scale precision calibration tasks at the LHC.
\end{abstract}

\hypersetup{%
pdfauthor={Fabrice Couderc},%
pdftitle={Lepton scales and smearing},%
pdfsubject={Calibration},%
pdfkeywords={Collider experiments, lepton calibration, photon calibration}} % limit six total

\maketitle

%%%%%%%% introduction %%%%%%%%
\section{Introduction}

A precise calibration of the photon and lepton energy scales is a key requirement for precision measurements at high-energy collider experiments. In particular, the measurement of the Higgs boson mass in decay channels involving photons or charged leptons, such as \(H \rightarrow \gamma\gamma\) and \(H \rightarrow ZZ^{*} \rightarrow 4\ell\), is directly sensitive to small biases in the energy or momentum scale, which needs to be known with a precision better than $0.1 \%$. Since electrons and photons share a largely common electromagnetic response in the detector, the energy-scale calibration of electrons is propagated to photons, with additional corrections accounting for differences in shower development and upstream material effects.

Decays of the \(Z\) boson into charged-lepton pairs, \zll with \(\ell\equiv e, \mu\), provide a primary in-situ reference for lepton energy or momentum calibration, owing to their large production rate and the precisely known \(Z\)-boson mass~\cite{PDG:2022}. The calibration is typically performed by constraining the reconstructed dilepton invariant-mass distribution to a reference line shape. However, the invariant mass depends simultaneously on the response of both leptons and is affected by the natural width of the \(Z\) boson, final-state radiation, and detector resolution effects. These features introduce correlations between scale and resolution parameters, complicating their simultaneous extraction in differential calibration schemes.

The calibration strategy is based on a comparison between the lepton-energy scale measured in data and that predicted by a detailed detector simulation. Residual differences between data and simulation are interpreted as corrections to the energy scale and resolution and are extracted using \zll decays assuming that biases on the lepton angular properties are negligible, which is usually the case in collider experiments. Within the framework we propose, these residual mismodeling effects are assumed to induce approximately Gaussian distortions of the di-lepton invariant-mass distribution, referred to as smearing hereafter.

The core of the method consists of describing the smearing effects using an analytical approach, enabling the construction of a fully analytic likelihood. Automatic differentiation techniques~\cite{autodiff} are then used to efficiently compute exact gradients with respect to all calibration parameters.

This method is implemented in a software tool named \ijazz\ (I Just AnalyZe the Z), which is freely available via a PyPI distribution~\cite{ijazz} and is described in section~\ref{sec:sas}. The treatment of statistical uncertainties arising from the finite size of the simulated \dyee sample is discussed in section~\ref{sec:mc_uncertainty}. Additional considerations related to energy-response linearity are presented in section~\ref{sec:linearity}, and a validation of the method with real LHC data is provided in section~\ref{sec:validation}. An extension of the method to photon energy-scale measurements using \dymmg\ decays is proposed in section~\ref{sec:method_mmg}.

\section{The \ijazz method}
\label{sec:sas}

The aim of \ijazz is to measure the differences between Monte Carlo (MC) simulation and data in terms of lepton energy response and resolution. These calibration parameters are usually measured with respect to a set of variables~$\vec{X}$ describing the lepton properties and detector conditions, such as the polar and azimuthal angles $\theta$ and $\phi$, the pseudorapidity $\eta\equiv -\ln[\tan(\theta/2)]$, and shower-shape observables characterizing the lateral and longitudinal development of the electromagnetic cluster in the calorimeter. The transverse momentum $p_{T}$ is omitted for the moment as it requires a dedicated treatment as it is already part of the invariant-mass calculation.

The di-lepton invariant mass is defined as:
\begin{equation}
m_{\ell\ell}^{2} = 2 p_{T,1} p_{T,2} \left[ \cosh(\Delta\eta) - \cos(\Delta\phi) \right],
\label{eq:mll}
\end{equation}
where $\Delta\eta$ and $\Delta\phi$ denote the differences in pseudorapidity and azimuthal angle between the two leptons, and $p_{T,1}$ and $p_{T,2}$ are their transverse momenta. The angular quantities $\Delta\eta$ and $\Delta\phi$ are typically extremely well measured in collider experiments; therefore, in the present approach, they are assumed to be perfectly known and not to introduce any bias or smearing in the reconstructed invariant mass.

We note $\rl(\vec X)$ and $\sl(\vec X)$, respectively the data/MC relative energy scale and data/MC energy smearing. To be more specific, in this method, we correct the energy from the simulation to match the one in data. Therefore, $\rl(\vec X)$ is a correction to be applied to the lepton energy in the simulation:
\begin{equation}
    E_\mathrm{corr}^{\mc} = \rl(\vec X) \times E_\mathrm{raw}^\mc
    \label{eq:mc_corr1}
\end{equation}
where $E_\mathrm{corr}^\mc$ is the scale-corrected energy while $E_\mathrm{raw}^\mc$ is the original lepton energy. Usually and conversely, one can correct the lepton energy in the data back to the simulation level with the formula:
\begin{equation}
    E_\mathrm{corr}^{\data} = E_\mathrm{raw}^\data / \rl(\vec X) \,.
    \label{eq:mc_corr_dt}
\end{equation}
Concerning the energy resolution, it is assumed to be always better (\ie smaller) in the simulation, thus the energy resolution in the simulation needs to be degraded (smeared) to its corresponding level in data.  Because the simulation already includes most of the effects due to the detector response (energy loss, material...), the modest degradation due to the imperfect modelling of the simulation is assumed to follow a normal distribution. The lepton-energy degradation is done in the simulation with random number trials from a normal distribution. Therefore, for each lepton in the simulation, its energy is drawn from the probabilistic distribution (including scale and smearing corrections):
\begin{equation}
    \mathcal{P}_\mathrm{corr}^\mc\!\left(E;\, \rl,\, \sl\right)  = \mathcal{N}\!\left(E;\ \rl(\vec X)\,E_\mathrm{raw}^\mc ,\
                        \rl(\vec X)\,\sl(\vec X)\times E_\mathrm{raw}^\mc \ \right),
    \label{eq:mc_corr}
\end{equation}
where $\mathcal{N}(x; \mu, \sigma)$ denotes a normal distribution with mean $\mu$ and standard deviation $\sigma$.

\subsection{Definition of the lepton categories and DY regions}

In order to extract the values of $\rl(\vec{X})$ and $\sl(\vec{X})$, the phase space defined by the variables $\vec{X}$ is first discretized into bins. For simplicity, a one-dimensional binning with $N_B$ bins is considered, each bin being denoted by $b$. The detector response and smearing are measured independently in each of these bins, resulting in $N_B$ scale and smearing parameters, collected in the vectors $\vec{\rl}$ and $\vec{\sl}$.

Since the $Z$ boson decays into a pair of leptons, the calibration must be propagated to the di-lepton invariant-mass scale and smearing, denoted by $\rll$ and $\sll$, respectively. These quantities are inferred from the single-lepton parameters $\rl$ and $\sl$ associated with the two leptons forming the $Z$-boson candidate.

Drell--Yan (DY) events are therefore categorized according to the pair of bins $(b_1, b_2)$ corresponding to the two leptons in the event, where $b_i$ denotes the response bin of lepton $i$. Since the configurations $(b_1, b_2)$ and $(b_2, b_1)$ lead to identical values of $\rll$ and $\sll$, they are assigned to the same category. The total number of independent di-lepton categories is thus
\begin{equation}
N_C \equiv \frac{N_B \times (N_B + 1)}{2}.
\end{equation}
The per-event di-lepton scale and smearing parameters, $\rll_c$ and $\sll_c$, associated with each category $c = (b_1, b_2)$ can be computed from Eq.~\ref{eq:mll} and the law of propagation of uncertainty. They are given by:

\begin{equation}
    \begin{split}
        \rll_c &= \sqrt{\rl_{b_1}\times\rl_{b_2}} \\
        \sll_c &= 0.5 \times \sqrt{\sl_{b_1}^2 + \sl_{b_2}^2}
    \end{split}
    \label{eq:sas_per_cat}
\end{equation}

\subsection{Definition of the likelihood}

In each di-lepton category $c$, the parameters $\rll_c$ and $\sll_c$ are determined by comparing the expected scaled and smeared di-lepton invariant-mass distribution predicted by simulation with the corresponding distribution observed in data. The comparison is performed by discretizing the di-lepton invariant-mass spectrum into $N_I$ bins. Although the number of bins $N_I$ may depend on the category $c$, this dependence is omitted for simplicity and this binning is denoted by $b_I$ in the following.

The comparison between data and simulation is carried out under the assumption that the predicted scaled and smeared distribution follows a multinomial probability distribution. Consequently, for each di-lepton mass bin $i$ in a given category $c$, a multinomial-based likelihood term is computed:
\begin{equation}
    \mathcal{L}(\nic; \rl_b, \sl_b) = \prod_{c = 1}^{N_C} \prod_{i=1}^{N_I} \frac{\left(\sum_i \nic\right)!}{\prod_i \nic!}\ \pic^{\nic}\ .
    \label{eq:sas_likelihood}
\end{equation}
where \pic is the expected probability (depending on $\rll_c$ and $\sll_c$ and consequently on $\rl_b$ and $\sl_b$) for a di-lepton event to fall in bin $i$ when belonging to category $c$ and \nic is the observed number of events in data corresponding to the same bin $i$. By definition we have $\sum_i \pic = 1$.
The likelihood from Eq.~\ref{eq:sas_likelihood} can be maximized with respect to the parameters $\rl_b$ and $\sl_b$. In practice, we minimize the negative log-likelihood, defined by:
\begin{equation}
    \nll(\nic; \rl_b, \sl_b) = - \sum_{c = 1}^{N_C} \sum_{i=1}^{N_I} \nic \log\left(\pic\right)\ ,
    \label{eq:nll}
\end{equation}
Additive constants from the multinomial coefficient in Eq.~\ref{eq:sas_likelihood} are omitted, as they are independent of the calibration parameters.

\subsection{Analytical scale and smearing corrections}
As a consequence of Eqs.~\ref{eq:mc_corr} and~\ref{eq:sas_per_cat}, each simulated di-lepton event with invariant mass $m_{\ell\ell}^{\mc}$ gives rise to a probabilistic scaled and smeared mass distribution, $\mathcal{M}_{\ell\ell}$, defined as
\begin{equation}
  \mathcal{M}_{\ell\ell}\!\left(m;\,\rll, \sll\right) =
  \mathcal{N}\!\left(
    m;\,\rll\, m_{\ell\ell}^{\mc},\, \rll\,\sll\, m_{\ell\ell}^{\mc}\ .
  \right),
  \label{eq:mll_gaus}
\end{equation}
Since this probabilistic model is compared to data using a binned invariant-mass distribution, the probability $M_i$ for a simulated event to fall into bin $i$ can be predicted as:

\begin{equation}
    M_i\!\left(m;\,\rll, \sll\right) = \frac{1}{2} \left[\mathrm{Erf}\left(\frac{b_i^\mathrm{u}/\rll - \mll^\mc}{\sqrt{2}\,\sll\,\mll^\mc}\right) -
                                                         \mathrm{Erf}\left(\frac{b_i^\mathrm{d}/\rll - \mll^\mc}{\sqrt{2}\,\sll\,\mll^\mc}\right)\right]
    \label{eq:ana_smearing}
\end{equation}
where $\mathrm{Erf}$ is the error function, $b_i^\mathrm{u}$ and $b_i^\mathrm{d}$ are respectively the upper and lower bound of bin $i$ of the di-lepton invariant mass distribution ($i\in[1, N_I]$).

Finally, the invariant-mass distribution in the simulation is binned with a very fine granularity in order to accelerate the computation of the per-event probability $\pic$. To avoid any bias, this binning, denoted $b^\mc_J$, must be smaller than the expected smearing $\sll$. For $\sll\approx~1.4~\gev$, several bin widths were tested ranging from $0.1~\gev$ to $0.5~\gev$ and no impact was observed on the fit results; in practice, a bin width below $0.2~\gev$ is recommended. The bins $b^\mc_J$ are indexed by $j$, and the number of simulated events falling into bin $j$ of category $c$ is denoted $\mjc$.

Considering Eq.~\ref{eq:ana_smearing}, the tensor $\aijc$ can be used to predict the contribution from bin $j$ of the simulation to bin $i$ in the likelihood computation. The tensor $\aijc$ is defined as:
\begin{equation}
    \aijc(\rll_c, \sll_c) = \frac{1}{2} \left[\mathrm{Erf}\left(\frac{b_i^\mathrm{u}/\rll_c - b^\mc_j}{\sqrt{2}\,\sll_c\,b^\mc_j}\right) -
                                              \mathrm{Erf}\left(\frac{b_i^\mathrm{d}/\rll_c - b^\mc_j}{\sqrt{2}\,\sll_c\,b^\mc_j}\right)\right]
    \label{eq:aijc}
\end{equation}
where $b^\mc_j$ is the center of bin $j$ of the binning $b^\mc_J$. Thus, we can predict the \pic probabilities with:
\begin{equation}
    \pic = \frac{\sum_j \aijc\ \mjc}{\sum_{i,j} \aijc\ \mjc}\ .
    \label{eq:pic}
\end{equation}
In Eq.~\ref{eq:pic}, the normalization factor in the denominator accounts for the fact that the fine binning $b^\mc_J$ can, and should, span a wider di-lepton invariant-mass range than the target binning $b_I$, in order to properly account for event migration into and out of $b_I$ due to the scaling and smearing applied to the simulation.

\begin{figure}[hbt]
    \centering
    \includegraphics[width=0.9\linewidth]{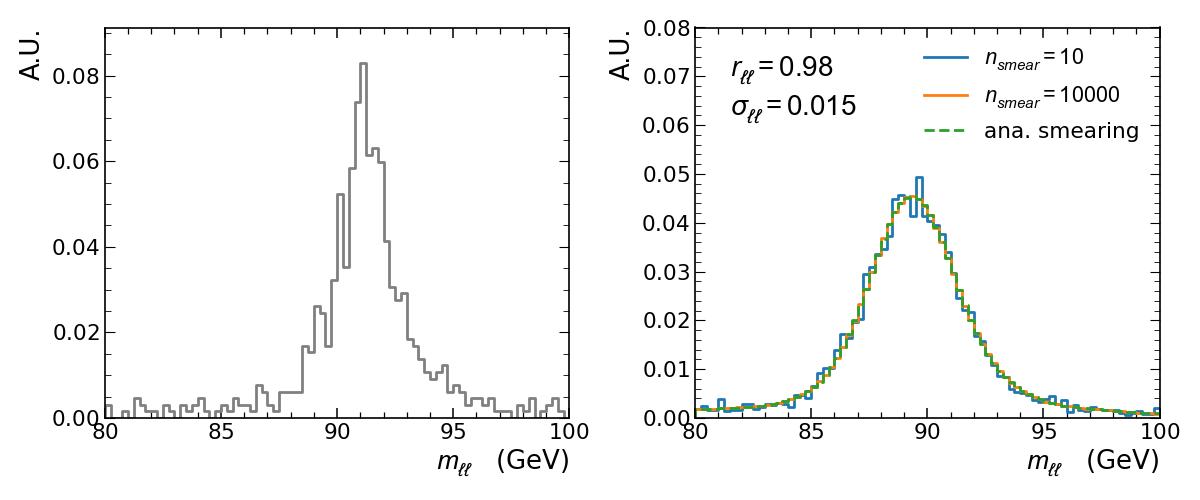}
    \caption{
    Left: the original MC distribution (Breit--Wigner) from 2,000 generated events.
    Right: the smeared MC distribution obtained using a random smearing technique, with $n_\mathrm{smear}$ trials per original MC event, together with the analytical smearing prediction (dashed line).
    It can be seen that the analytical smearing accurately reproduces the expected distribution, which is approximated by $n_\mathrm{smear}=10{,}000$; the two distributions are indistinguishable.
    }
    \label{fig:ana_smear}
\end{figure}

The validity of the method is illustrated in Fig.~\ref{fig:ana_smear}, which compares a Breit--Wigner distribution of 2,000 events that have been scaled and smeared with $\rll=0.98$ and $\sll=0.005$. For each simulated event, a varying number of random trials was generated and compared to the analytical prediction. As the number of trials increases, the agreement improves; with a large number of trials, $n_\mathrm{smear} \geq 10{,}000$, the analytical prediction and the Monte Carlo simulation are in perfect agreement.

\subsection{Qualitative comparison: analytical vs random smearing techniques}

Per-event probabilities $\pic$ can be computed using a random smearing method with typically $n_\mathrm{smear}=10$ trials per simulated event (\eg, see Ref.~\cite{CMS:egm}). The analytical smearing method presented here provides a CPU-time gain of $\sim 500$ for $10^6$ MC events, up to $\sim 5000$ for $10^7$ events on a laptop with 6 CPU cores.

This gain is further enhanced by automatic-gradient computations implemented via TensorFlow~\cite{tensorflow}. With random smearing, the likelihood gradient must be computed numerically, requiring the likelihood to be evaluated twice per fit parameter. Thus the time $t_\mathrm{num}$ needed to compute the numerical gradient is $t_\mathrm{num} \propto 2 \times N_\mathrm{par}$, where $N_\mathrm{par}$ is the number of fit parameters. In contrast, for the analytical method, the gradient evaluation is roughly independent of the number of parameters, representing an additional gain of time for large $N_\mathrm{par}$. Overall, the total CPU-time gain for likelihood maximization is several orders of magnitude, and can be further accelerated on GPUs. For example, a minimization with 100 parameters and $2\times 10^7$ events takes 20--30~min on a laptop, less than one minute on a GPU, whereas the same fit with random smearing would take several days.

\subsection{Treatment of statistical uncertainties}
\label{sec:stat_err}

The uncertainties arising from the finite size of the data sample are computed using the covariance matrix, $\Sigma$, defined as the inverse of the Hessian of the negative log-likelihood, $H^\nll$:
\begin{equation}
    \Sigma^{-1}_{kp} \equiv H^\nll_{kp} =
    \frac{\partial^2 \nll\!\left(\nic;\, \vec{\theta}\right)}{\partial \theta_k \, \partial \theta_p},
    \label{eq:hessian}
\end{equation}
where $\vec{\theta}$ denotes the ensemble of parameters $\rl_b$ and $\sl_b$. Note that this uncertainty does not account for statistical fluctuations due to the finite size of the Monte Carlo simulation, which in some cases may be comparable to that of the data sample.

\subsection{Validation with a naive MC approach}
\label{sec:validation1}

The method is first validated by generating $25\times 10^6$ MC events according to a Cauchy distribution with mean $\mu = \MZ$ and width $\gamma = \Gamma_Z/2$, where $\MZ$ and $\Gamma_Z$ are the mass and natural width of the $Z$ boson, respectively, taken from Ref.~\cite{PDG:2022}. In this validation, the full kinematics of the two leptons is not simulated; only a naive $Z$-boson line shape is generated using a standard Breit--Wigner distribution.

For each event, two random numbers are generated to mimic a property $X$ of the two leptons from the $Z$-boson decay, with $X$ drawn from a uniform distribution over the range $[0, 100]$. The detector energy resolution is simulated by generating, for each event, a random number from a normal distribution with mean 1 and standard deviation 0.015. For a subset of $5\times 10^6$ events, an additional miscalibration and smearing depending on $X$ is applied, emulating the difference between the true detector response (data) and the nominal simulation (MC). The \ijazz method is then used to retrieve these miscalibration and smearing parameters, treating the latter part of the sample (with the additional miscalibration) as "data" and the rest as "MC".

\begin{figure}[hbt]
    \centering
    \includegraphics[width=0.9\linewidth]{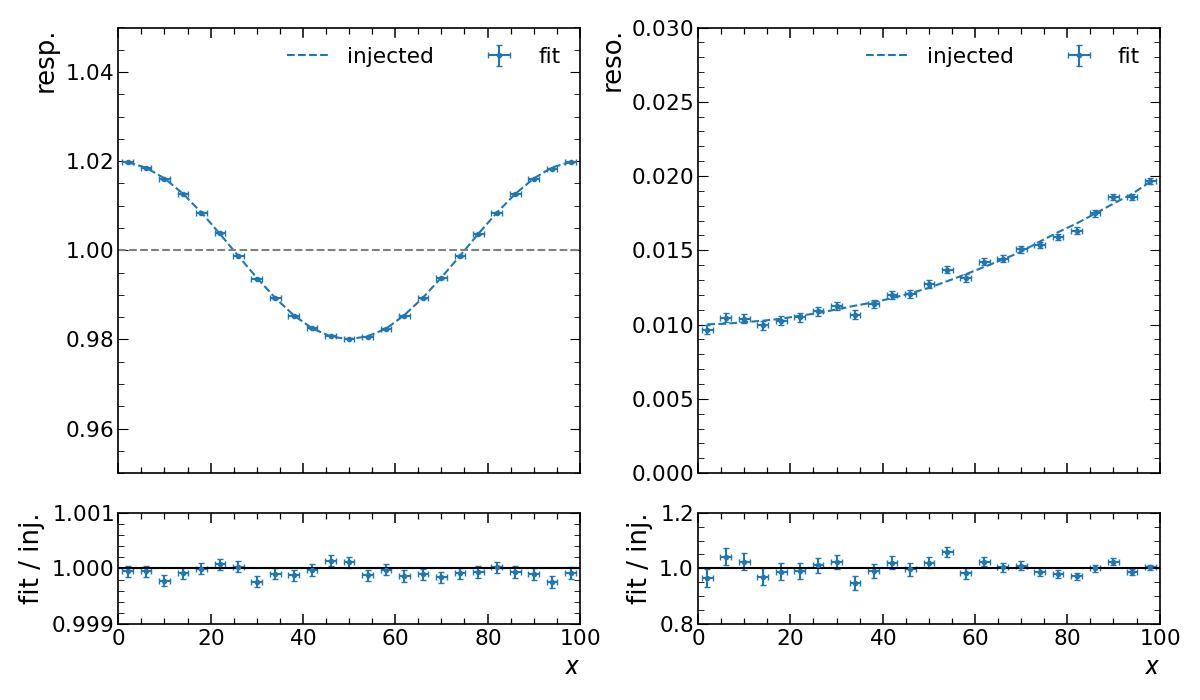}
    \caption{
    Validation of the method using a naive MC simulation. A subset of events is miscalibrated and smeared according to known functions (injected curves). These parameters are accurately recovered by the fit (points). The top panels show the absolute values of the parameters, while the bottom panels display the ratio of fitted to injected values. The left (right) plots correspond to the response ($\rl_b$) and resolution ($\sl_b$) parameters, respectively.
    }
    \label{fig:validation1}
\end{figure}

The results of this validation are shown in Fig.~\ref{fig:validation1}, demonstrating that the fitted parameters agree with the injected ones within statistical precision.

\subsection{Adaptive binning \texorpdfstring{$b_I$}{bI}}
Because the predicted \pic enter the \nll function through $\log\pic$, zero probabilities must be avoided. In addition, the invariant-mass spectrum is sharply peaked, and low-statistics tails can bias the fit (\eg by favoring artificially large smearing to populate empty bins). To mitigate this, we use an adaptive binning to determine $b_I$: for each category, bin edges are defined so that each mass bin contains the same number of events. In addition, a minimum number of MC events is required per category (100 events by default). The total number of bins in $b_I$ is inspired by the Freedman--Diaconis rule~\cite{FreedmanDiaconis} and is taken to be the minimum between $\left(\sum_i \nic\right)^\frac{1}{3}$ and $\Delta M_{\ell\ell}/(0.5~\gev)$ with $\Delta M_{\ell\ell}$ the size of the di-lepton mass window over which the fit is performed, typically $\Delta M_{\ell\ell} = (100-80)~\gev$, the limit on the total number of bins in $b_I$ comes from the fact that the typical energy-resolution is larger than $0.5~\gev$ at the $Z$-boson peak. This default behaviour is configurable.

\begin{figure}[htb]
    \centering
    \includegraphics[width=0.9\linewidth]{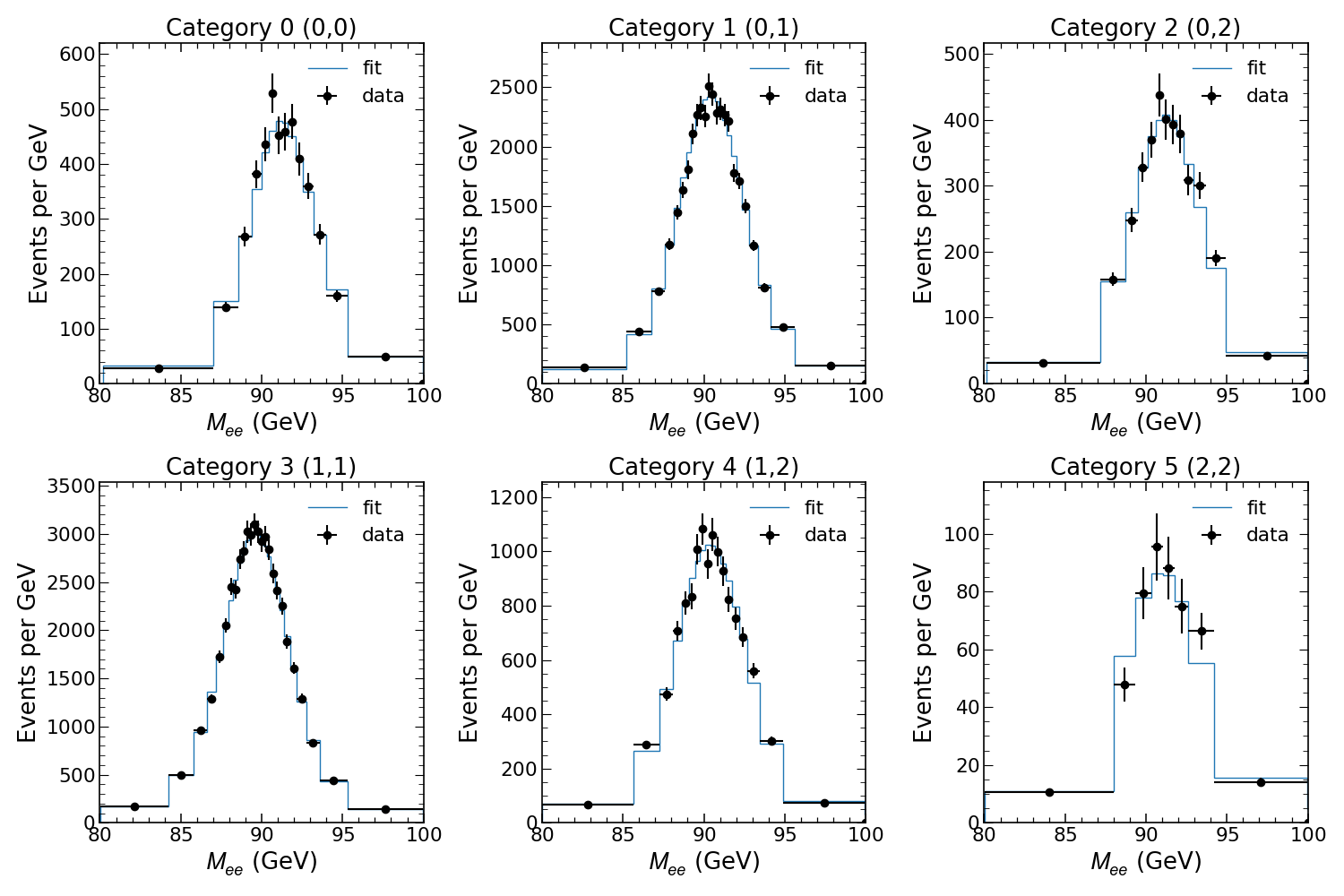}
    \caption{adaptive binning: The bin width is adapted so that the total number of events in the simulation is the same in each bin. The total number of bins in each category depends on the available MC statistics in that category, \emph{e.g.}, the bottom-right category has substantially lower statistics, leading to a binning with a larger overall bin width than the other categories. The points correspond to the fitted data (in this case a toy MC dataset) while the histogram refers to the fitted prediction from the smeared simulation.}
    %For instance the bottom right category has a much smaller statistics yielding to a binning with an overall bin width larger than the other categories.
    \label{fig:adaptive_binning}
\end{figure}

A demonstration of this technique is presented on Fig.~\ref{fig:adaptive_binning}. A similar simulation as the one described in section~\ref{sec:validation1} is used but we reduce the total number of events to $5\times10^4$ for both data and simulation and the property $X$ is generated according to a normal distribution so as to obtain some categories with lower statistics, the number of scale and smearing parameters is $N_B = 3$, consequently giving rise to $N_C = 6$ categories.

%%%%%%% mc uncertainty %%%%%%%%
\section{Uncertainties due to limited MC statistics}
\label{sec:mc_uncertainty}
As discussed in Sec.~\ref{sec:stat_err}, the covariance matrix obtained from the method only accounts for statistical fluctuations due to the finite size of the data sample, assuming that the predicted probabilities $\pic$ are computed with infinite precision. In reality, $\pic$ are limited by the finite size of the simulation sample. To include this effect, one can study the variation of the \nll minimum induced by fluctuations of the simulated counts $\mjc$.

If $\mjc$ fluctuates by $\djc$, the negative log-likelihood becomes $\nll_\delta$, which can be written as
\begin{equation}
    \nll_\delta(\vtheta, \djc) = \nll(\vtheta) + \delta \nll(\vtheta, \djc),
\end{equation}
where $\vtheta$ collectively denotes the response and resolution parameters of the \nll function. To first order in $\djc$,
\begin{equation}
    \delta \nll(\vtheta, \djc) = \frac{\partial \nll}{\partial \mjc}(\vtheta) \, \djc
    = \left[\sum_i \frac{n_c}{p_c} \aijc - \sum_i \frac{\nic}{\pic} \aijc \right] \djc,
\end{equation}
with $n_c \equiv \sum_i \nic$ and $p_c \equiv \sum_i \pic$, the second line being obtained by differentiating Eq.~\ref{eq:nll} with respect to $\mjc$.

Expanding $\nll_\delta$ to second order in $\delta \vtheta \equiv \vtheta - \vthetast$, one obtains
\begin{equation}
    \nll_\delta(\vtheta, \djc) = \nll(\vthetast)
    + \frac{\partial \nll}{\partial \vtheta}(\vthetast)\, \delta \vtheta
    + \frac{1}{2}\, \delta \vtheta^T H_\nll \, \delta \vtheta
    + \frac{\partial \nll}{\partial \mjc}(\vthetast) \, \djc
    + \frac{\partial^2 \nll}{\partial \vtheta \, \partial \mjc}(\vthetast) \, \delta \vtheta \, \djc,
    \label{eq:nll_delta_2dorder}
\end{equation}
where $\vthetast$ denotes the \nll minimum. Differentiating Eq.~\ref{eq:nll_delta_2dorder} with respect to $\vtheta$ gives the minimum condition:
\begin{equation}
    H_\nll \, \delta \vtheta + \frac{\partial^2 \nll}{\partial \vtheta \, \partial \mjc}(\vthetast) \, \djc = 0,
\end{equation}
which implies
\begin{equation}
    \delta \vtheta_{jc} = - H_\nll^{-1} \, \frac{\partial^2 \nll}{\partial \vtheta \, \partial \mjc}(\vthetast) \, \djc.
\end{equation}

Since the $\djc$ fluctuations are independent, the total uncertainty on $\vthetast$ due to limited MC statistics is obtained by summing in quadrature:
\begin{equation}
    (\delta \vthetast)^2 = \sum_{j,c} (\delta \vtheta_{jc})^2
    = \sum_{j,c} \left[ H_\nll^{-1} \, \frac{\partial^2 \nll}{\partial \vtheta \, \partial \mjc}(\vthetast) \right]^2 \, \djc^2.
    \label{eq:mc_err}
\end{equation}

For a weighted simulation, $\mjc = \sum_q w_{jcq}$, where $w_{jcq}$ is the weight of event $q$, the corresponding fluctuation is $\djc = \sqrt{\sum_q w_{jcq}^2}$.

\begin{figure}[htb]
    \centering
    \includegraphics[width=0.9\linewidth]{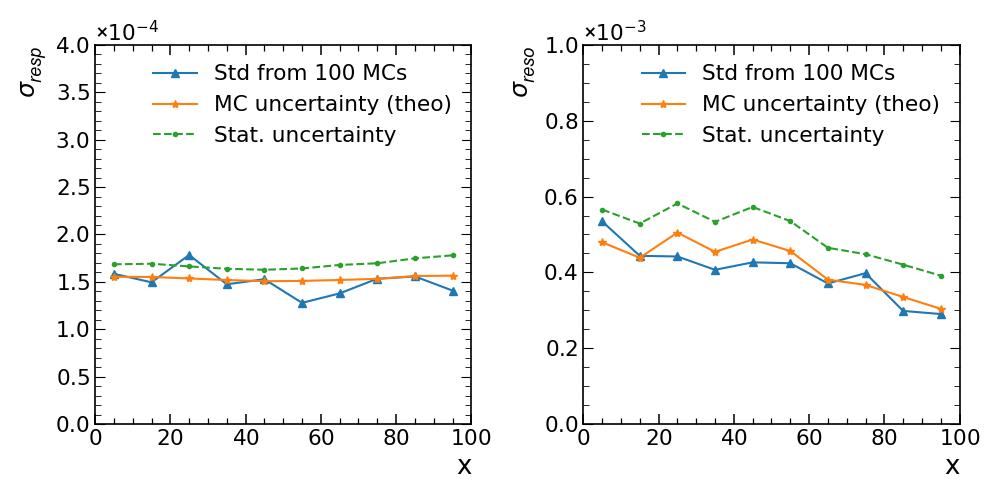}
    \caption{
    Validation of the statistical uncertainties due to limited MC statistics for the response parameters (left) and smearing parameters (right). Triangles show the standard deviation of 100 measurements using the same data but different simulations, while stars show the MC uncertainty predicted by Eq.~\ref{eq:mc_err}. The dashed line indicates the statistical uncertainty from the data alone; since data and MC sample sizes are equal, MC and data uncertainties are of the same order.
    }
    \label{fig:mc_err}
\end{figure}

To validate this formula, we use the simple simulation described in Sec.~\ref{sec:validation1}: $2\times 10^6$ events are treated as the data sample, and the scale and smearing parameters are measured for 100 independent simulations of $2\times 10^6$ events each. The standard deviation of these 100 measurements represents the uncertainty due to MC fluctuations. Figure~\ref{fig:mc_err} compares these standard deviations to the predictions of Eq.~\ref{eq:mc_err}, showing very good agreement.

%\section{categorization vs \texorpdfstring{\pt}{pT} and validation with a Pythia-based MC simulation}
\section{Energy non-linearity treatment via \texorpdfstring{\pt}{pT} categorization and validation with a Pythia-based MC simulation}

\label{sec:linearity}
The analytical smearing formalism does not by itself solve category-migration biases when the categorization variable is correlated with the fitted observable. Therefore, when the lepton energy or transverse momentum \pt is included in the set of properties $\vec{X}$, a special treatment is required. In such cases, \pt is used for categorization, so the corrected \pt may fall into a different category than the original one, leading to category migrations and correlations between the categorization and the reconstructed di-lepton invariant mass. Since for a quasi-massless particle, its transverse momentum \pt is directly proportional to its energy $E$ \emph{via} $E=p_T\times\cosh\eta$, at fixed $\eta$, binning the energy-scale correction factor in \pt and $\eta$ is fully equivalent to binning in energy and $\eta$, thereby accounting for the full energy non‑linearity. Because of the invariance of the lepton \pt spectrum under a Lorentz boost along the beam axis in \dyll events, we recommend using a \pt, $\eta$ binning to measure the energy non-linearity.

Since the biases related to the non-linearity of energy response are strongly correlated with the lepton \pt spectrum, we validate the method using a realistic DY simulation generated with Pythia~\cite{pythia83}. For each lepton from the $Z$-boson decay, the energy is first smeared according to a normal distribution with mean $\mu_\mathrm{sim} = 1$ and standard deviation $\sigma_\mathrm{sim} = 1.5\%$, representing the typical energy resolution of the ATLAS or CMS electromagnetic calorimeters. The resulting event properties are denoted as $sim$: ${\mll}_\mathrm{sim}$, ${p_T}_\mathrm{sim}$, etc. A total of $30\times 10^6$ events are generated in this way. Half of the sample is used as reference simulation, while the other half is additionally miscalibrated and smeared according to known (injected) parameters to emulate the difference between the true detector response and the nominal simulation.

\begin{figure}[htb]
    \centering
    \includegraphics[width=0.9\linewidth]{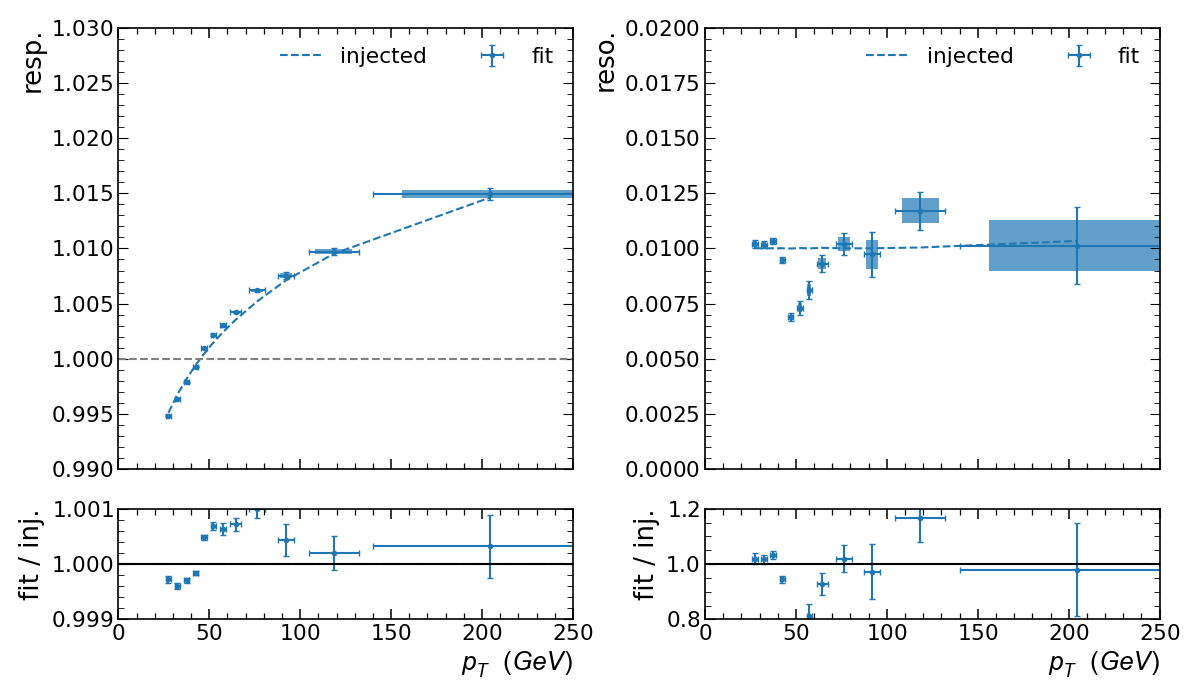}
    \caption{
    Measured scale (left) and smearing (right) parameters using the Pythia-based DY simulation. The dashed line corresponds to the injected parameters, while the shaded bands indicate the contribution from limited MC statistics to the total uncertainty. A \pt-dependent categorization introduces small biases in the measured scale and larger biases in the smearing parameters.
    }
    \label{fig:pt_sas}
\end{figure}

We then measure the scale and smearing parameters using a lepton-\pt dependent categorization. The results are shown in Fig.~\ref{fig:pt_sas}. A small bias is observed in the measured scale parameters (below $0.1\%$), while a larger bias affects the smearing parameters, consistent with expectations from category migration effects.

\begin{figure}[htb]
    \centering
    \includegraphics[width=0.9\linewidth]{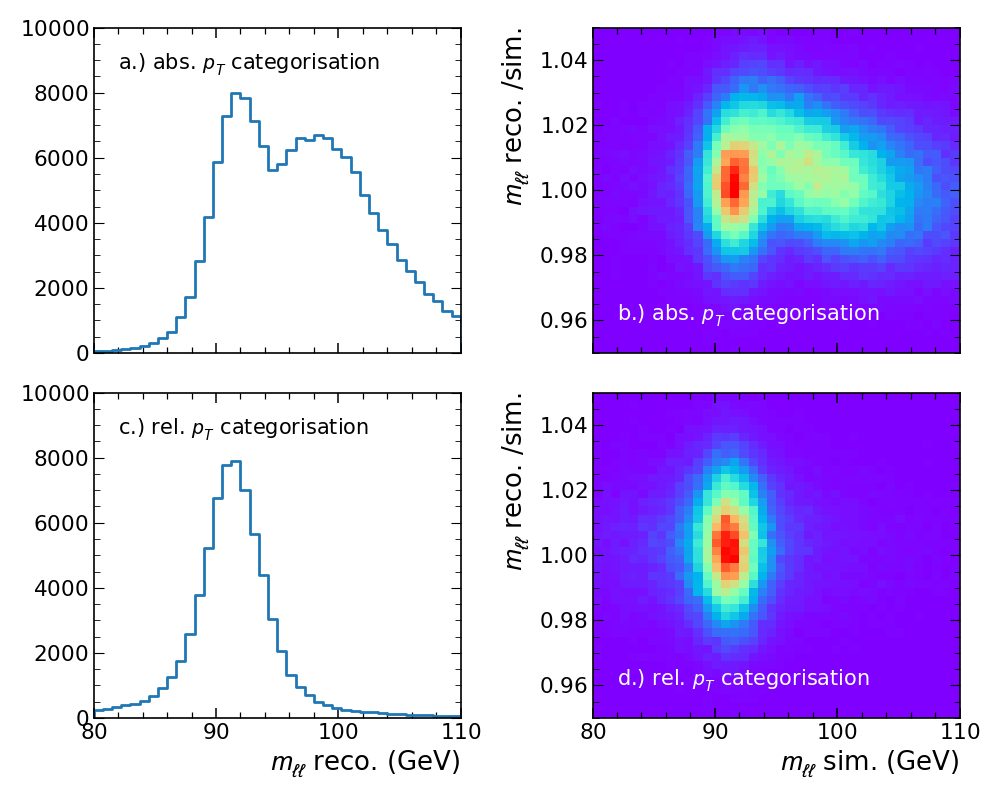}
    \caption{
    Left: reconstructed di-lepton mass for the absolute \pt selection (top) and relative \pt selection (bottom). Right: correlation between the reconstructed-to-simulated di-lepton mass ratio ($\mll/{\mll}_\mathrm{sim}$) and the simulated mass. Using the relative \pt selection removes the double-peak structure and associated correlations observed with the absolute \pt selection.
    }
    \label{fig:mee_sim_reco}
\end{figure}

To better understand these biases, we select events in which one lepton satisfies $45~\gev< {\pt}_\ell < 50~\gev$ and the other $50~\gev< {p_T}_\ell < 60~\gev$. The reconstructed di-lepton invariant-mass distribution, $\mll$, for these events exhibits a double-peak structure, as shown in Fig.~\ref{fig:mee_sim_reco}a. This arises from a threshold effect induced by the \pt requirements. Furthermore, the two peaks show different correlations with the original simulated mass, ${\mll}_\mathrm{sim}$: the peak near the $Z$-boson mass is largely uncorrelated, whereas the secondary peak exhibits a strong correlation. These correlations are illustrated in Fig.~\ref{fig:mee_sim_reco}b. When the two peaks mix (for instance, for symmetric \pt requirements), biases in the measured scale and smearing parameters can arise.

These biases can be mitigated by selecting events based on the relative transverse momentum, defined as $r_{\pt} = {\pt}_\ell / \mll$. Using this criterion instead of the absolute \pt, we select one lepton with $45/\MZ < {\pt}_\ell / \mll < 50/\MZ$ and the other with $50/\MZ < {\pt}_\ell / \mll < 60/\MZ$, where $\MZ$ is the $Z$-boson mass~\cite{PDG:2022}. Figs.~\ref{fig:mee_sim_reco}c and \ref{fig:mee_sim_reco}d show the reconstructed $\mll$ distribution and its correlation with the simulated mass under this selection, demonstrating that the double-peak structure and the related correlations disappear with the relative \pt requirement.

In light of the properties of the relative \pt categorization, we adopt the following procedure to measure the scale and smearing parameters. The lepton-\pt categorization is replaced by a relative \pt categorization using, instead of \pt, the variable $r_{\pt} = \pt / \mll$. The relative \pt binning is defined by dividing the original \pt bin edges $[{\pt}_1, {\pt}_2, \dots, {\pt}_n]$ by the $Z$-boson mass, $\MZ$.

A first fit is performed to extract the energy-scale parameters, after which the lepton \pt and the di-lepton mass $\mll$ are corrected using Eq.~\ref{eq:mc_corr_dt}. A second fit is then performed to measure the energy-smearing parameters. Indeed, the relative \pt categorization introduces correlations between the two leptons, so the scale must be set to 1 in order to properly extract the smearing parameters.

However, in order to correct the scale or smear the lepton energies in the simulation, an absolute lepton \pt criterion is required. Therefore, the relative \pt binning is transformed back into an absolute \pt binning. In each relative \pt bin $b$ of $b_I$, the average absolute transverse momentum, ${\ptbar}_{b}$, is computed and used to define the absolute \pt bin edges as
\[
[p_{T1}, 0.5({\ptbar}_{1}+{\ptbar}_{2}), \dots, 0.5({\ptbar}_{b_k}+{\ptbar}_{b_{k+1}}), \dots, {\pt}_n].
\]
It is observed that this recasted \pt binning closely reproduces the original \pt categories used as input.

\begin{figure}[htb]
    \centering
    \includegraphics[width=0.9\linewidth]{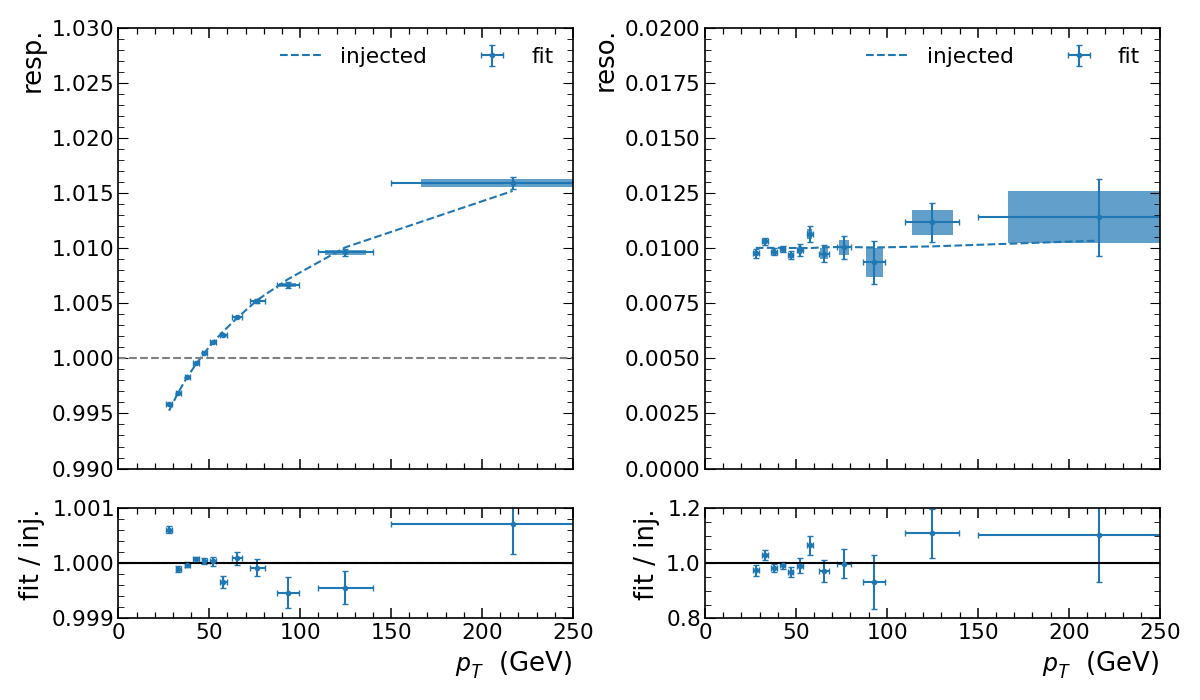}
    \caption{
    Measured scale (left) and smearing (right) parameters using the relative \pt categorization with the Pythia-based DY simulation described in the text. The dashed line corresponds to the injected parameters, and the shaded bands indicate the contribution of limited MC statistics to the total uncertainty.
    }
    \label{fig:rpt_sas}
\end{figure}

The results of this procedure are shown in Fig.~\ref{fig:rpt_sas}, where the $x$-axis corresponds to the recasted absolute \pt binning. The injected parameters shown in the figure are measured in the recasted \pt categories, meaning they correspond to the scale and smearing parameters that would be applied to correct the lepton energies in data/simulation.

Note that this two-step fit procedure does not improve the smearing parameters in the scenario of Fig.~\ref{fig:pt_sas}. For the absolute \pt binning case, the second fit converges to the same smearing parameters as the first one, as expected, because no correlations are introduced between the leptons, unlike in the relative \pt case.

From Fig.~\ref{fig:rpt_sas}, one can see that both the scale and smearing parameters are correctly retrieved within uncertainties, with the exception of the first \pt bin for the scale parameter, which is slightly biased ($< 0.1~\%$) due to the absolute \pt cut, $p_T > 25~\gev$. If deemed relevant, the first \pt bin should either be discarded in the analysis, or an additional bin should be added at lower \pt to absorb the effect of the threshold.

\section{Validation with LHC real data}
\label{sec:validation}
\begin{figure}[htb]
    \centering
    \includegraphics[width=0.45\linewidth]{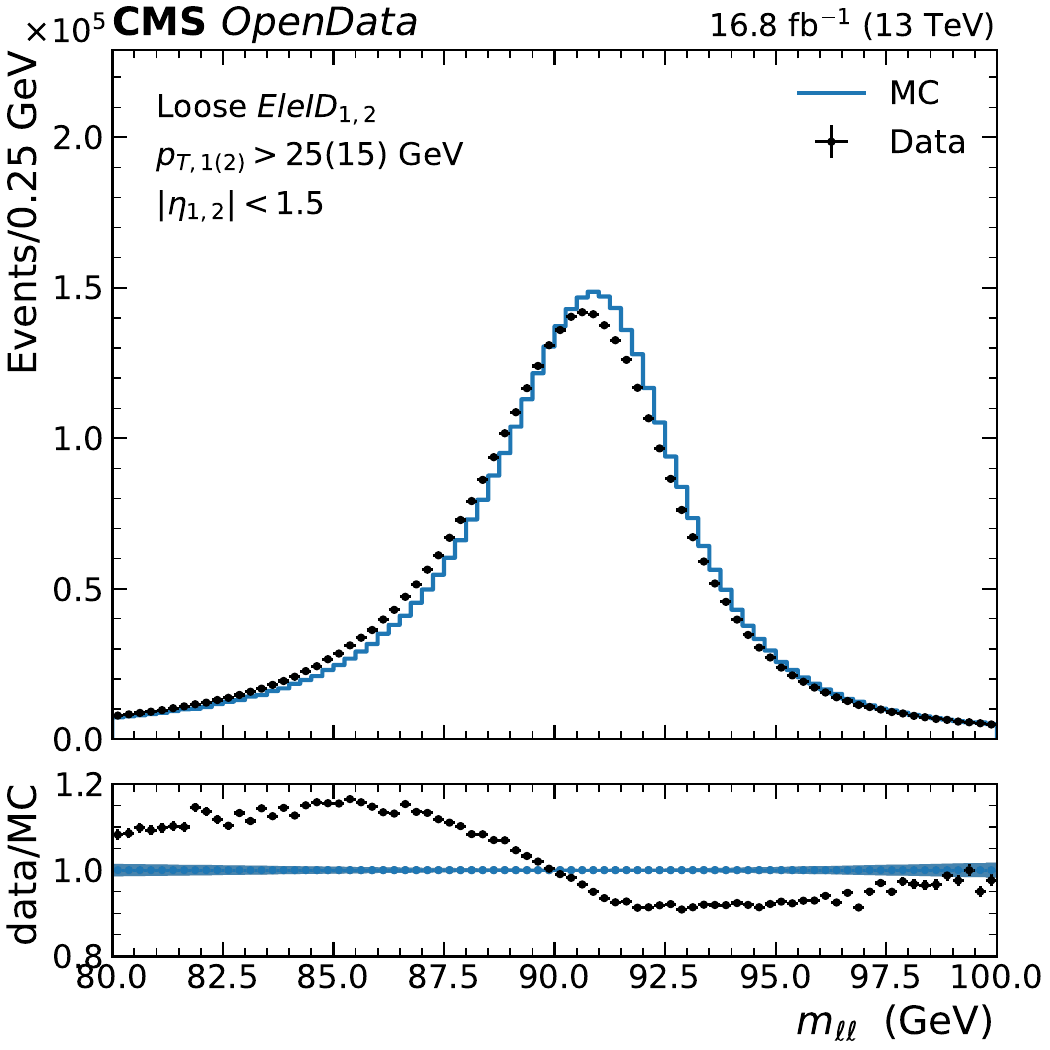}
    \includegraphics[width=0.45\linewidth]{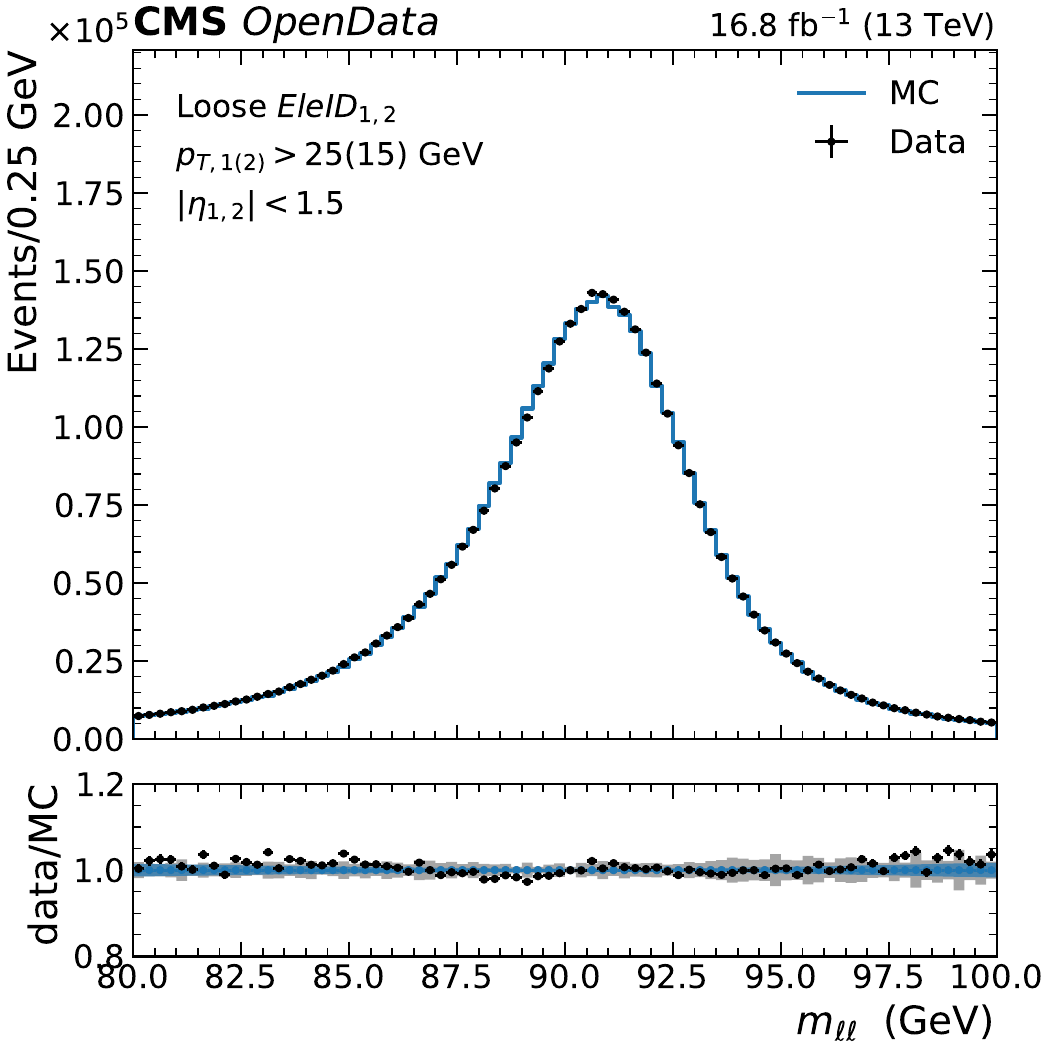}
    \caption{
    Validation of the \ijazz method using CMS Open Data. Left (resp. right): DY invariant mass without (resp. with) the \ijazz corrections. The blue-line histogram represents the simulation while the points with error bars correspond to the data. The gray bands in the lower panels show the total systematic uncertainty (see text) while the blue bands refer to the MC simulation statistical uncertainty only.
    }
    \label{fig:validation_sas}
\end{figure}

In order to validate the method with real LHC data, we apply the \ijazz fit to \dyee decays recorded by the CMS experiment in 2016~\cite{CMS_DoubleEG_2016}. Data are compared to predictions~\cite{CMS_DYJetsToLL_2016} based on events from the MC@NLO~\cite{mcatnlo} generator passed through a complete GEANT4~\cite{geant4} simulation of the CMS detector. In this example, the energy scale and smearing corrections are computed in two stages using a simplified set of variables that adequately describe electromagnetic (EM) showers in CMS: $\eta$, $R_9$ and \pt, where $R_9$ represents the lateral extension of the EM shower in the CMS calorimeter~\cite{CMS:egm}. First, the energy scale is equalized with respect to $\eta$ and $R_9$ using the binning: $\eta \in [-2.5, -2.0, -1.5, -1.2, -1.0, 0.0, +1.0, +1.2, +1.5, +2.0, +2.5]$ and $R_9\in [0, 0.96, 1.0]$. Then, the energy-scale linearity is corrected using the binning $|\eta| \in [0.0, 1.0, 1.5, 2.0, 2.5]$ and $p_T\in[0, 25, 35, 40, 50, 65, 80, 100, +\infty]~\gev$. The energy-scale corrections are ranging between a few per mille at $\eta\approx 0$ to the percent level in the endcaps, similarly the MC smearing parameter $\sl$ ranges from $0.8~\%$ at $\eta\approx 0$ to a few percent at $\eta\gtrsim 2.0$. The data/MC comparisons of the Z-boson invariant mass lineshape are shown in Fig.~\ref{fig:validation_sas} before and after the energy scale and smearing corrections. In the lower panels of the figure, the data/MC ratio is given along with the MC statistical uncertainty (blue band) and the systematic uncertainty (gray band). To compute the systematic uncertainties, we vary a condition, redo the fit, and take the difference with respect to the nominal fit as uncertainty. In this study, we have included a limited set of systematic uncertainties for demonstration purposes:
\begin{itemize}
    \item Fit window: this accounts for the effect of potential non‑Gaussian tails in the assumption used in Eq.~\ref{eq:mll_gaus}. We vary the fit window from $[80, 100]~\gev$ (nominal fit) to $[75, 105]~\gev$ (systematic fit).
    \item Electron identification: to account for potential additional EM shower shape dependence, we vary the identification of the electron (EleID) used in the fit, since the identification discriminators are correlated with EM showers. We vary the EleID criteria from the CMS loose EleID (nominal fit) to the CMS tight EleID (systematic fit). See~\cite{CMS:egm} for more information on the EleID criteria.
\end{itemize}

Despite the simplified approach, both in terms of the binning choices and the treatment of systematic uncertainties, an excellent agreement between data and simulation is already observed. For a more complete calibration study, the analyst should consider a broader set of experimental effects affecting the dilepton invariant-mass distribution, including pileup dependence, FSR modelling, background contamination, trigger and reconstruction efficiencies, residual mismodelling of the Z-boson kinematics\dots

\section{Extension to photon energy scale measurement}
\label{sec:method_mmg}

The method developed for lepton energy scale and smearing determination can be further extended to measure the photon energy scale and smearing in \dymmg decays.

Following the definition of the scale and smearing parameters introduced in section~\ref{sec:sas}, we denote $\rg(\vec X)$ and $\sg(\vec X)$ as the data-to-simulation relative photon-energy scale and the data-to-simulation photon-energy smearing, respectively. Similarly to Eq.~\ref{eq:mc_corr_dt}, the photon energy in data can be corrected back to the simulation level by:
\begin{equation}
    E_\mathrm{corr}^{\data} = E_\mathrm{raw}^\data / \rg(\vec X)\,.
    \label{eq:mc_corrg_dt}
\end{equation}

The method requires specific adaptations for the \dymmg process. On the one hand, the presence of a single photon in the final state simplifies the formulation; on the other hand, the photon carries only a fraction of the total energy of the three-body system. To account for this feature, we introduce a new variable, $V_\dy$, representing the relative mass carried by the photon in the \dymmg decay. This variable replaces the di-lepton invariant mass in the analytical likelihood, allowing the same fitting procedure as for the lepton case. The method is validated using a Pythia-based simulation.

\subsection{Adapting the method to the \texorpdfstring{\dymmg}{Zmmg} process}
\label{sec:method:description}

The squared invariant mass of the three-body system in the \dymmg process can be written as
\begin{equation}
\Mmmg^2 = \Mmm^2 + 2\,\ptg\left(
    \ptmum \,(\cosh\Delta\eta_{\gamma\mum} - \cos\Delta\phi_{\gamma\mum}) +
    \ptmup \,(\cosh\Delta\eta_{\gamma\mup} - \cos\Delta\phi_{\gamma\mup})
    \right),
\label{eq:invmass_mmg}
\end{equation}
where $\ptmum$, $\ptmup$, and $\ptg$ are the transverse momenta of the two muons and the photon, respectively, and $\Mmmg$ and $\Mmm$ denote the invariant masses of the $\mup\mum\gamma$ and $\mup\mum$ systems. Considering a variation $\delta\ptg$ of the photon transverse momentum, the three-body invariant mass can be expressed as
\begin{equation}
    \Mmmg\left(\delta \ptg\right) =
    \sqrt{\Mmm^2 + \left(1 + \frac{\delta\ptg}{\ptg}\right)\left(\Mmmg^2 - \Mmm^2\right)}\,.
    \label{eq:invmass_mmmg_dr}
\end{equation}
Assuming that $\delta\ptg/\ptg \ll 1$, the corresponding variation of the invariant mass, $\delta \Mmmg\equiv \Mmmg(\delta \ptg) - \Mmmg$, is given by
\begin{equation}
    \frac{\delta \Mmmg}{\Mmmg} =
    \frac{1}{2}\frac{\delta\ptg}{\ptg}\left(1 - \frac{\Mmm^2}{\Mmmg^2}\right).
\end{equation}
In the following, we define $\drg \equiv \delta\ptg/\ptg$, such that $\rg = 1 + \drg$.

Assuming that the two muons are well calibrated and that the photon energy scale is close to its nominal value (for instance, using electromagnetic calibrations derived from \dyee events), one has $\Mmmg \approx \MZ$, where $\MZ$ is the $Z$-boson pole mass. In this case,
\[
\frac{\delta \Mmmg}{\Mmmg} \approx \frac{\Mmmg}{\MZ} - 1\,.
\]
Combining the above expressions, we define
\begin{equation}
\Vdy \equiv \left(\frac{\Mmmg}{\MZ} - 1\right)
\frac{2}{1 - \frac{\Mmm^2}{\Mmmg^2}},
\label{eq:vdy}
\end{equation}
which is directly sensitive to residual miscalibration of the photon energy. A relative photon-energy scale miscalibration $\delta\rg \ll 1$ therefore shifts the \Vdy distribution by approximately $\rg - 1$, yielding
\begin{equation}
\Vdy(\rg) \approx \Vdy(\rg = 1) + (\rg - 1)\,.
\label{eq:vdy_sas}
\end{equation}

\begin{figure}[htb]
    \centering
    \includegraphics[width=0.70\textwidth]{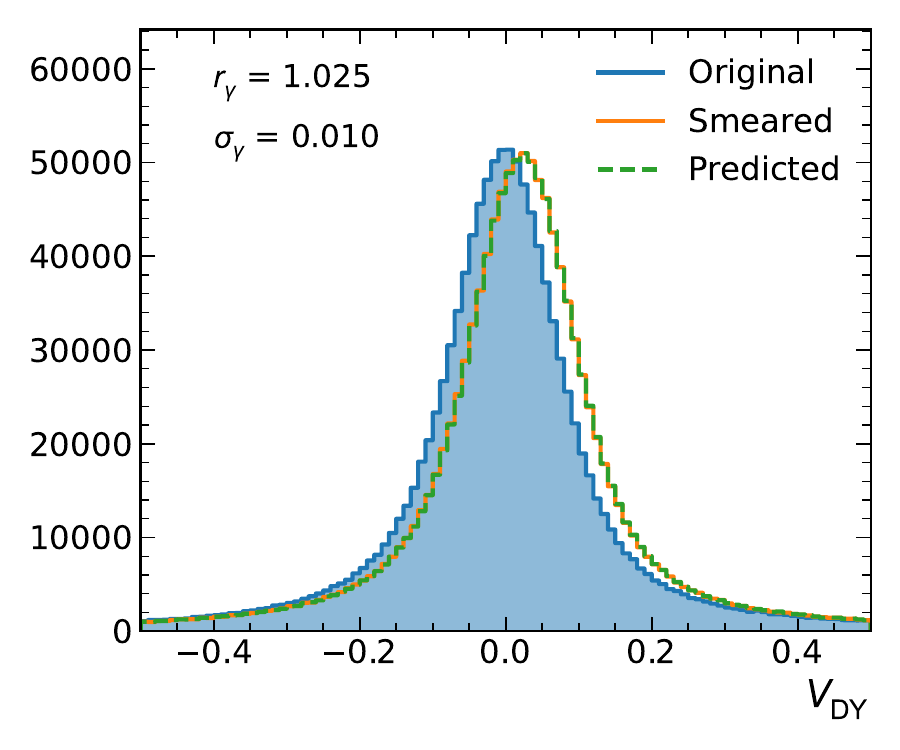}
    \caption{
    Validation of Eq.~\ref{eq:vdy_sas} using a Pythia-based simulation of the \dymmg process. The original distribution represents \Vdy in the original simulation. The smeared distribution corresponds to \Vdy after applying the photon energy shift and smearing described in the text. The predicted distribution is obtained from Eq.~\ref{eq:vdy_sas}, starting from the original distribution. The smeared and predicted distributions are nearly indistinguishable.
    }
    \label{fig:vdy_sas}
\end{figure}

Figure~\ref{fig:vdy_sas} illustrates a validation of this relation using a Pythia-based simulation of the \dymmg process, selecting events with $\ptg > 25~\gev$. For each event, the photon energy is rescaled according to $\rg = 1 + \drg$, where $\drg$ is a random number drawn from a normal distribution $\mathcal{N}(\drgm, (1+\drgm)\sg)$. We use the value $\drgm = 0.025$ which corresponds to a $2.5\%$ energy shift, while $\sg = 0.01$ mimics a degradation of the photon-energy resolution. The predicted \Vdy distribution obtained from Eq.~\ref{eq:vdy_sas} is found to be in excellent agreement with the simulation.

\subsection{Photon analytical likelihood}
\label{sec:method:smearing}

The analytical likelihood method developed for the lepton case can be directly applied to the \Vdy distributions. For each simulated event, Eq.~\ref{eq:vdy_sas} is used to generate the scaled and smeared distribution of $\Vdy^\mc$:
\begin{equation}
    \mathcal{P}_\mathrm{corr}^{\mc}\!\left(v;\, \rg,\, \sg \right)
    = \mathcal{N}\!\left(v;\, \Vdy^\mc + \rg - 1,\; \rg \times \sg \right).
    \label{eq:method_mmg}
\end{equation}

Accordingly, the \aijc tensor defined in Eq.~\ref{eq:aijc} becomes
\begin{equation}
\aijc(\rg_c, \sg_c) =
\frac{1}{2}\Bigg[
\mathrm{Erf}\!\left(
\frac{b_i^\mathrm{u} - b_j^\mc - (\rg_c - 1)}
     {\sqrt{2}\,\rg_c\,\sg_c}
\right)
-
\mathrm{Erf}\!\left(
\frac{b_i^\mathrm{d} - b_j^\mc - (\rg_c - 1)}
     {\sqrt{2}\,\rg_c\,\sg_c}
\right)
\Bigg],
\label{eq:aijc_mmg}
\end{equation}
where $b_i^\mathrm{u,d}$ denote the upper and lower bin edges of the \Vdy distribution in data, $b_j^\mc$ are the bin centers in simulation, and $c$ labels the event category. This tensor predicts the contribution of each simulation bin $b_j^\mc$ to the predicted probability $\pic$ in bin $i$ of Eq.~\ref{eq:nll},
\begin{equation}
\pic =
\frac{\sum_j \aijc\, {\Vdy}_{jc}}
     {\sum_{i,j} \aijc\, {\Vdy}_{jc}}\,.
\label{eq:pic_mmg}
\end{equation}

The rest of the analytical likelihood procedure remains unchanged with respect to the lepton case, with the only modification being the replacement of the invariant-mass distributions by the \Vdy distributions.

\subsection{Validation using Pythia-based simulations}
\label{sec:method:validation}

The method is validated using a Pythia-based simulation of the $pp \to Z/\gamma^* \to \mup\mum\gamma$ process at $\sqrt{s} = 13.6$~\tev. Typical event selections used in collider experiments are emulated by applying the following requirements: $|\eta_{\mu^{\pm}}| < 2.5$, $|\eta_\gamma| < 2.5$, ${\pt}_\mu^{\pm} > 15$~\gev and $\ptg>25$~\gev. After these selections, the total simulated sample contains approximately $30 \times 10^6$ events. Several tests are performed by varying the injected photon-energy miscalibration parameters. For each test, the full sample is split into two subsamples: one is used to represent the data, while the other serves as the reference simulation. In each event of the data subsample, the photon energy is miscalibrated by $\rg = 1 + \delta\rg$, where $\delta\rg$ is a random number drawn from a normal distribution $\mathcal{N}(\drgm, (1+\drgm)\sg)$. Unless otherwise stated, a value of $\sg = 1\%$ is used.

Since Eq.~\ref{eq:vdy_sas} is only an approximation, an iterative procedure is required to retrieve the injected miscalibration. From Eqs.~\ref{eq:vdy} and~\ref{eq:invmass_mmmg_dr}, the first-order Taylor expansion of $\Vdy$ in $\drg$ is given by:
\begin{equation}
    \begin{split}
    \Vdy(\delta\rg) &= \Vdy(0) + (1-\epsilon)\,\delta\rg,\ \mathrm{with} \\
    \epsilon &\equiv 1 - \left.\frac{\partial \Vdy}{\partial \delta\rg}\right|_{\delta\rg=0}
    = \left(1 - \frac{\Mmmg}{\MZ}\right)\times\frac{3\kmm - 2}{\kmm}
    \end{split}
\end{equation}
where $\kmm \equiv 1 - \Mmm^2/\Mmmg^2$. Since $\Mmmg / \MZ \approx 1$, the parameter $\epsilon$ is small, with $|\epsilon| < 0.10$ in practice. Because Eq.~\ref{eq:method_mmg} is used to smear the simulation, an iterative fitting procedure is required. After the first iteration, the fitted value converges to ${\delta\rg}_\mathrm{fit}^1 = (1 - \epsilon)\,\drgm$. More generally, one can show recursively that after iteration $n$, the fitted value is
\[
{\delta\rg}_\mathrm{fit}^n = \left(1 - \epsilon^n\right)\,\drgm .
\]
The fit therefore converges rapidly to the true value $\drgm$, typically within a few iterations.

\begin{figure}
    \centering
    \includegraphics[width=0.98\textwidth]{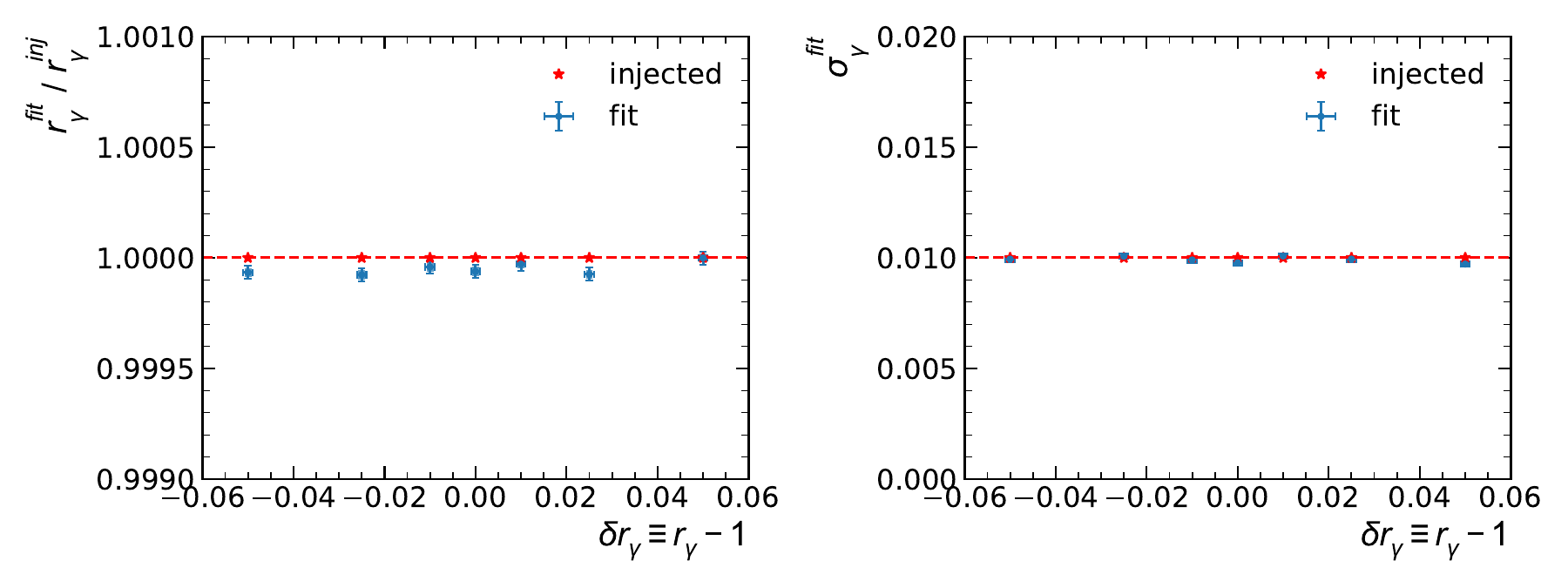}
    \caption{Validation of the method using a Pythia-based \dymmg simulation. Left: retrieved value of the photon energy scale shift \drg as a function of the injected \drgm value. Right: retrieved value of the photon energy resolution smearing $\sg$ as a function of the injected \drgm value. The error bars represent the total statistical uncertainties, including both data and simulation contributions, the latter being shown by the solid blue boxes. The star markers indicate the injected values.}
    \label{fig:validation_sas_pythia}
\end{figure}

The convergence is illustrated in Fig.~\ref{fig:validation_sas_pythia}, where $\drgm$ is varied between $-5\%$ and $+5\%$. The fit is performed in the invariant-mass range $80~\gev < \Mmmg < 100~\gev$. The method accurately retrieves the injected values of both $\drg$ and $\sg$. The accuracy of the $\drgm$ determination is better than $10^{-4}$. A small residual bias is nevertheless observed in Fig.~\ref{fig:validation_sas_pythia}; its origin is discussed in Section~\ref{sec:spurious_shift}. This result demonstrates the validity of the full method, including the iterative fitting procedure.

\subsection{Intrinsic limitations of the method}
If the difference in energy resolution between simulation and data is large, two potential biases may affect the energy scale measurement. These effects are discussed in this section.

\subsubsection{Spurious energy shift arising from energy-resolution mismodeling}
\label{sec:spurious_shift}

The second-order Taylor expansion of Eq.~\ref{eq:vdy} can be written as:
\begin{equation}
    \Vdy(\delta\rg) = \Vdy(0) + (1-\epsilon)\,\delta\rg
    + \frac{1}{2}
    \left.\frac{\partial^2 \Vdy}{\partial \delta\rg^2}\right|_{\delta\rg=0}
    (\delta\rg)^2 \, .
    \label{eq:vdy_2dorder}
\end{equation}

In this method, the miscalibration $\delta\rg$ between data and simulation is assumed to follow a normal distribution. If we assume that there is no energy-shift between the data and the simulation, \ie when the mean value of $\delta\rg$ vanishes: $\langle\drg \rangle = \drgm = 0$, a shift may arise between the mean value $\langle \Vdy \rangle_{\data}$ measured in data and its corresponding value $\langle \Vdy \rangle_{\mc}$ predicted by the simulation due to the presence of the second-order term. By definition, for a centered normal distribution, one has $\langle (\delta\rg)^2 \rangle = \sg^2$. Therefore, in this case, Eq.~\ref{eq:vdy_2dorder} transforms in:
\begin{equation}
    \langle \Vdy \rangle_{\data}
    =
    \langle \Vdy \rangle_{\mc}
    + \frac{1}{2}
    \left.\frac{\partial^2 \Vdy}{\partial \delta\rg^2}\right|_{\delta\rg=0}
    \sg^2 \, .
    \label{eq:sg_shift1}
\end{equation}

The second-order term reads:
\begin{equation}
    \frac{1}{2}
    \left.\frac{\partial^2 \Vdy}{\partial \delta\rg^2}\right|_{\delta\rg=0}
    =
    2\,\frac{1-\kmm}{\kmm}
    \left(\frac{\Mmmg}{\MZ} - 1\right)
    - \left(1 - \frac{3}{4}\kmm\right)
    \frac{\Mmmg}{\MZ} \, .
    \label{eq:shift_2dorder}
\end{equation}
Thus, even if the data are perfectly calibrated, the fit may assign an apparent energy shift of
\(
{\delta\rg}_{\mathrm{fit}}
\approx
\frac{1}{2}
\left.\frac{\partial^2 \Vdy}{\partial \delta\rg^2}\right|_{\delta\rg=0}
\sg^2 .
\)
Since $\Mmmg / \MZ \approx 1$, Eq.~\ref{eq:shift_2dorder} shows that this shift is always negative. The validity of Eq.~\ref{eq:shift_2dorder} has been verified directly using the \Vdy distributions.

\begin{figure}
    \centering
    \includegraphics[width=0.98\textwidth]{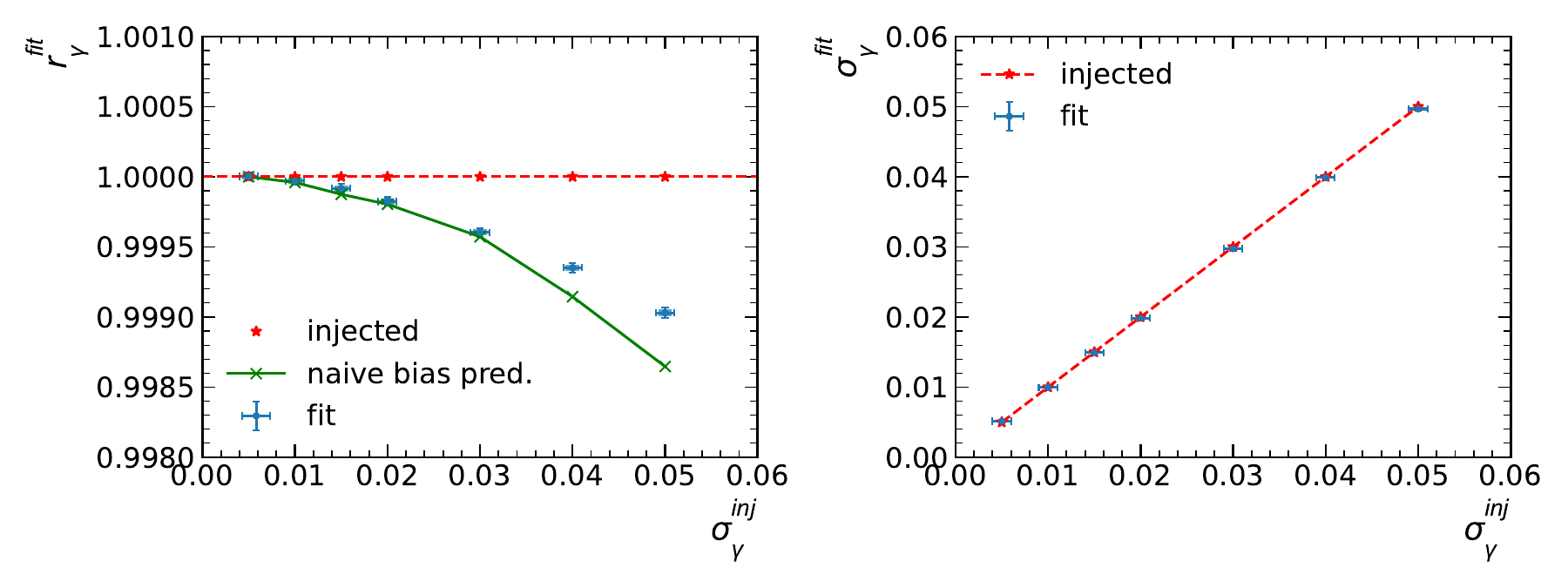}
    \caption{Potential bias induced by a difference in energy resolution between data and simulation. Left: retrieved value of the photon energy scale shift \drgm as a function of the injected $\sg$ value. A small bias is observed for large values of $\sg$. A naive prediction of this bias is also shown (see text), although it is not expected to perfectly reproduce the observed effect. Right: retrieved value of the photon energy resolution smearing $\sg$ as a function of the injected $\sg$ value. The error bars represent the total statistical uncertainties, including both data and simulation contributions, the latter being shown by the solid blue boxes. The star markers indicate the injected values.}
    \label{fig:sg_shift1}
\end{figure}

However, the impact on the fitted parameters differs from a simple shift of the mean, as it depends on the full shape of the \Vdy distribution. This effect is illustrated in Fig.~\ref{fig:sg_shift1}, where a Pythia-based simulation is used without any injected energy-scale shift between data and simulation, \ie $\drgm =0$, while only the value of $\sg$ is varied. The prediction of the resulting bias is not straightforward; a naive estimate is shown in Fig.~\ref{fig:sg_shift1}. It is obtained by propagating the shift from Eq.~\ref{eq:sg_shift1} to the mean of the smeared \Vdy distribution using the expected probabilities $\pic$ from Eq.~\ref{eq:pic_mmg}.

Fig.~\ref{fig:sg_shift1} shows that this effect can be neglected for $\sg < 0.02$. If required, it can also be calibrated out.

\subsubsection{Biases arising from the \texorpdfstring{\ptg}{pTg} selection}
\def\ptgtrue{\ensuremath{{\ptg}_\mathrm{true}}\xspace}
\def\ptgmeas{\ensuremath{{\ptg}_\mathrm{meas}}\xspace}

In \dymmg decays, the photon transverse momentum follows a steeply falling spectrum. As a consequence, when a selection such as $\ptg > c_\gamma$ is applied, a bias may be introduced. We denote by \ptgtrue\ the true transverse momentum of the photon and by \ptgmeas\ the measured one. Due to the combination of the steeply falling spectrum and energy-resolution effects, more photons with $\ptgtrue < c_\gamma$ fluctuate into the selected region $\ptgmeas > c_\gamma$ than photons with $\ptgtrue > c_\gamma$ fluctuate out of it. This results in an excess of photons with $\drg > 0$ entering the selection compared to photons with $\drg< 0$ leaving it. The net effect is therefore a positive bias on $\drg$ in the selected sample.

\begin{figure}
    \centering
    \includegraphics[width=0.48\textwidth, trim=0 0 15cm 0, clip]{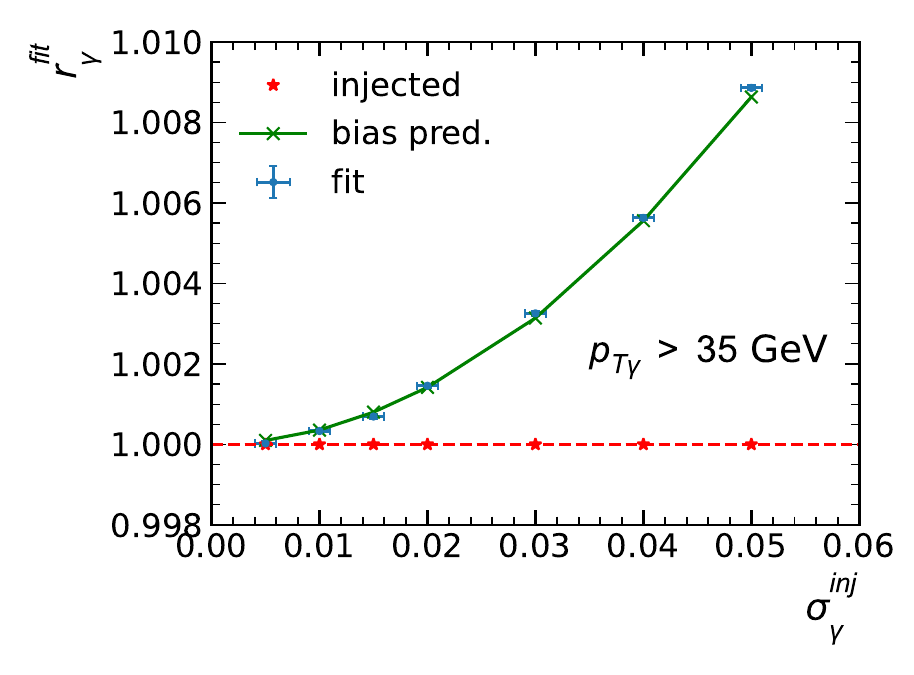}
    \includegraphics[width=0.48\textwidth, trim=0 0 15cm 0, clip]{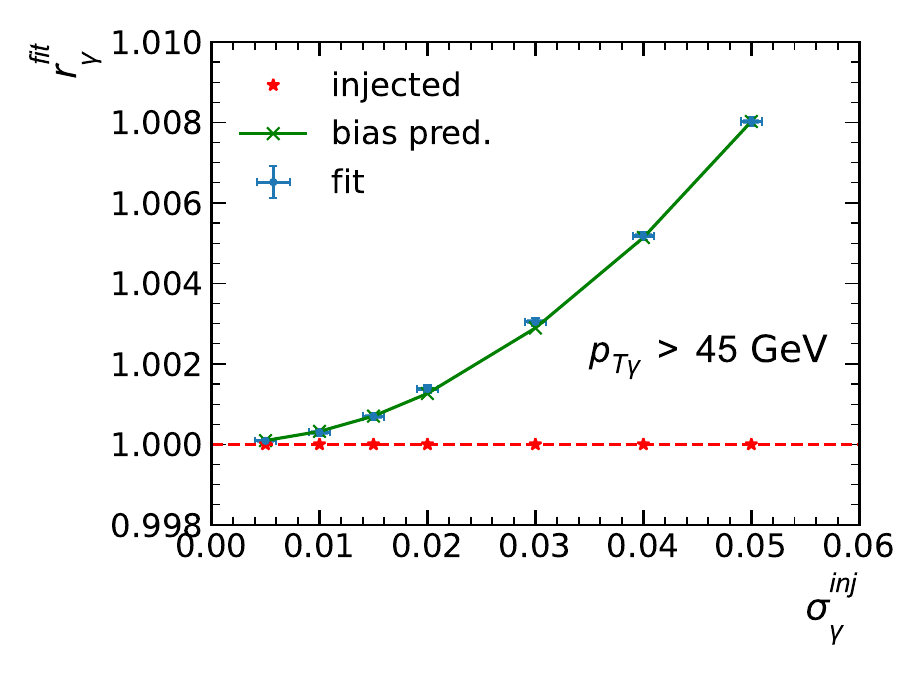}

    \caption{Potential bias induced by the steeply falling photon transverse-momentum spectrum combined with energy-resolution mismodeling.
    Left: retrieved value of the photon energy scale shift \drgm as a function of the injected $\sg$ value when applying the selection $\ptg > 35~\gev$.
    Right: retrieved value of the photon energy scale shift \drgm as a function of the injected $\sg$ value when applying the selection $\ptg > 45~\gev$.
    A bias is observed in the fitted values due to the \ptg selection, as described in the text. The predicted bias is also shown.
    The error bars represent the total statistical uncertainties, including both data and simulation contributions, the latter being shown by the solid blue boxes. The star markers indicate the injected values.}
    \label{fig:sg_shift2}
\end{figure}

To illustrate this effect, a Pythia-based simulation is used while varying the energy-resolution miscalibration parameter $\sg$. The data are assumed to be perfectly calibrated, \ie $\overline{\delta\rg} = 0$. A selection $\ptgmeas > c_\gamma$ is applied and the fit is performed. The results are shown in Fig.~\ref{fig:sg_shift2}. In this figure, the predicted energy bias corresponds to the sum of the negative bias discussed in section~\ref{sec:spurious_shift} and the positive bias induced by the \ptg selection, quantified as $\langle \delta\rg \rangle_\mathrm{cut}$ for photons passing the $c_\gamma$ requirement. An excellent agreement is observed between the predicted and the measured bias.

\subsubsection{Considerations on the limitations}

These limitations presented in this section are both related to the magnitude of the parameter $\sg$. In principle, they can be suppressed by smearing the simulation with the fitted value of $\sg$ and repeating the fit to extract the energy scale with an excellent precision. However, this procedure is not implemented in the current version of the software, and the analyst must carefully assess the origin of fitted value of $\sg$ before correcting this bias.

In practice, the smearing parameter can be precisely calibrated using \dyee decays with the method described in Section~\ref{sec:sas}, after which the photon energy in the simulation should be smeared accordingly. As a consequence, the parameter $\sg$ represents an uncertainty on the photon energy resolution which should be typically well below the percent. In this case, the overall effect on the photon energy scale is below $0.05~\%$ based on Fig.~\ref{fig:sg_shift2}. If the value of $\sg$ extracted from \dymmg decays is significant, it may indicate additional sources of smearing affecting the \Mmmg invariant-mass distribution, potentially related to muon momentum calibration and/or photon angular resolution effects. In such cases, the parameter $\sg$ effectively absorbs these mismodeling effects. This does not, however, introduce the biases discussed above, which are specifically associated with genuine photon energy-resolution mismodeling.

The interpretation of a large fitted value of \sg (\ie $>1~\%$) is therefore left to the analyst, who may choose either to correct for these effects by smearing the simulation prior to performing the fit, if necessary, or to assign a corresponding systematic uncertainty.

\section{Conclusion}

The method presented in this paper provides a new framework for the determination of lepton energy scale and resolution corrections based on an analytical treatment of detector smearing effects. By expressing the comparison between data and simulation through a fully differentiable likelihood, the approach eliminates random smearing techniques and enables an efficient use of automatic differentiation for likelihood minimization.

The method is implemented in the \ijazz software and validated using toy Monte Carlo studies and realistic Pythia-based DY simulations, demonstrating an accurate recovery of scale and resolution parameters within uncertainties. The impact of finite simulation statistics is consistently propagated to the fitted parameters through an analytical uncertainty estimate.

Lepton categorizations involving transverse momentum are shown to introduce biases due to kinematic correlations and category migration. A relative-\(p_T\) categorization strategy is proposed to mitigate these effects and to enable unbiased measurements in realistic conditions.

The method has been tested on real data from the LHC recorded by the CMS experiment, demonstrating an excellent data-to-MC agreement after the \ijazz corrections. Examples of systematic uncertainties are proposed, while it is left to the analyst to define the exhaustive list of potential systematic sources induced by any disagreement between data and simulation.

The approach is further extended to the determination of the photon energy scale using \dymmg events.

Thanks to its analytical formulation and compatibility with modern machine-learning frameworks, the proposed approach significantly improves numerical stability and reduces the computational cost by several orders of magnitude compared to traditional random smearing methods, making it well suited for large-scale precision calibration tasks at the LHC. The \ijazz software is freely available to the community via a PyPI distribution~\cite{ijazz} including relevant documentation and examples.

\begin{acknowledgments}
The authors are grateful to Nathalie Besson and Marco Pieri for their thorough reading of the manuscript and for insightful suggestions that significantly improved the clarity and quality of this document.
\end{acknowledgments}

\clearpage
\pagebreak

% \bibliography{biblio}

\end{document}